\documentclass{aa}  

\bibpunct{(}{)}{;}{a}{}{,} 

\usepackage{graphicx}
\usepackage{txfonts}
\usepackage{natbib,twoopt}
\usepackage[breaklinks=true]{hyperref} 
\bibpunct{(}{)}{;}{a}{}{,}             
\makeatletter
  \newcommandtwoopt{\citeads}[3][][]{\href{http://adsabs.harvard.edu/abs/#3}%
    {\def\hyper@linkstart##1##2{}%
     \let\hyper@linkend\@empty\citealp[#1][#2]{#3}}}
  \newcommandtwoopt{\citepads}[3][][]{\href{http://adsabs.harvard.edu/abs/#3}%
    {\def\hyper@linkstart##1##2{}%
     \let\hyper@linkend\@empty\citep[#1][#2]{#3}}}
  \newcommandtwoopt{\citetads}[3][][]{\href{http://adsabs.harvard.edu/abs/#3}%
    {\def\hyper@linkstart##1##2{}%
     \let\hyper@linkend\@empty\citet[#1][#2]{#3}}}
  \newcommandtwoopt{\citeyearads}[3][][]%
    {\href{http://adsabs.harvard.edu/abs/#3}
    {\def\hyper@linkstart##1##2{}%
     \let\hyper@linkend\@empty\citeyear[#1][#2]{#3}}}

\newcommand{\ang}{\AA ngstr{\"o}m }

\usepackage[]{color}

\makeatother

\hypersetup{draft}  
\begin{document}

   \title{Non-grey dimming events of KIC 8462852 from GTC spectrophotometry}

   \author{H. J. Deeg  \inst{1,2}
          \and  R. Alonso \inst{1,2}
          \and D. Nespral  \inst{1,2}
          \and{Tabetha Boyajian\inst{3}}
          }

   \institute{Instituto de Astrof\'\i sica de Canarias, C. Via Lactea S/N, E-38205 La Laguna, Tenerife, Spain
   \and
Universidad de La Laguna, Dept. de Astrof\'\i sica, E-38206 La Laguna, Tenerife, Spain\\
              \email{hdeeg@iac.es}
         \and
        Louisiana State University, Dept of Physics \& Astronomy, Baton Rouge, LA 70803, USA\\
             \email{}
             }

   \date{Received XXXX; accepted XXXX}

 
  \abstract
 {We report ground-based spectrophotometry of KIC 8462852, during its first dimming events since the end of the Kepler mission. The dimmings show a clear colour-signature, and are deeper in visual blue wavelengths than in red ones. The flux loss' wavelength dependency can be described with an \AA ngstr{\"o}m absorption coefficient of $2.19\pm0.45$, which is compatible with absorption by optically thin dust with particle sizes on the order of 0.0015 to 0.15 $\mu$m. These particles would be smaller than is required to be resistant against blow-out by radiation pressure when close to the star. {During occultation events, these particles must be replenished on time-scales of days.} If dust is indeed the source of KIC 8462852's dimming events, deeper dimming events should show more neutral colours, as is expected from optically thick absorbers.}   

   \keywords{stars: individual (KIC 8462852) — stars: peculiar }

   \maketitle
%

\section{Introduction}

The discovery of the strange brightness variations of the star KIC 8462852 or Boyajian's star (from here on denoted simply as 'KIC8462') in post-mission data from the Kepler satellite by \citetads{2016MNRAS.457.3988B} 
 has given rise to one of the most enigmatic astronomic discoveries in recent time. This star displayed two periods of strong dips in brightness, both lasting a few weeks and separated by about two years. It is a normal F star in most respects and quiescent most of the time, but has also been claimed to have undergone a 20\% brightness-loss since 1890 and the present \citepads{2016ApJ...822L..34S} 
, as well as a 0.34\% brightness loss per year during the 4 years of the Kepler mission, with some periods of stronger decrease \citepads{2016ApJ...830L..39M}. 
 Several families of explanations for its dimming events have been discussed by \citetads{2016ApJ...829L...3W} 
 , although none of them entirely satisfying. While most attention has been given to occultations from a swarm of very large comet-like objects \citepads{2016MNRAS.457.3988B,2016ApJ...819L..34B,2017arXiv171005929W}, 
 further recent efforts have focused on the explanation of KIC8462's brightness variations through intervening dust, arising from dust-enshrouded planetesimals \citepads{2017A&A...600A..86N} or circumstellar dust \citepads{2017ApJ...847..131M}. 

Since the strange behaviour of KIC8462 was not discovered until long after the end of the Kepler mission in 2013, the only data that had been available from its dimming events were the single-band ones from that mission. Arguing that colour-signatures during KIC8462's dimming events might be indicative of their nature, we obtained since the 2017A semester some target-of-opportunity observing time with the Gran Telescopio Canarias (GTC), in order to perform multicolour spectrophotometry of KIC8462 in the case that this target exhibits relevant new dimming events. Fortunately, during late Spring and Summer 2017, KIC8462 already began to exhibit several such events, with flux variations of up to 2\%, which we observed at 13 different occasions. Here we present these observations and their analysis with regard to the hypothesis of dusty occulters. These observations were also part of a larger community-wide follow-up program whose results are published by \citet[][furhter on B18]{B18}. 

 \section{Observations}
 Spectrophotometric observations of the target were obtained with the OSIRIS long-slit spectrograph using the R1000R grism, a slit-width of 12\arcsec\  and a 2 x 2 binning of the CCD. These data were obtained through the target-of-opportunity program 89-GTC54/17A, which was triggered for the first time on May 17, 2017 after receiving an alert of a possible variation in KIC8462's spectrum through the community monitoring program described in B18. This alert enabled us to commence with observations by GTC at the onset of the first known dimming event since the end of the Kepler mission.  Until Dec. 5, 2017, we acquired 16 pointings on KIC8462, covering several further dimming-events, as well as moments when the target had returned close to the brightness prior to these events (see Table 1 for dates). Unfortunately, a technical defect forced a shut-down of GTC for several weeks between mid-September and late November 2017, during which KIC8462 exhibited fluxes up to 1\% higher than previously recorded (see Fig.~\ref{fig:lco_gtc_phot}). The three GTC pointings in early December were taken when the target had nearly returned to the assumed quiescent flux. Each pointing consisted of a timeseries of about 30 min duration, based on a series of exposures with an open shutter time of 20 sec and cycle-time of about 45 sec (varying exposure times were used in the first pointing). A nearby star of similar brightness but redder colours (KIC 8462763, V=11.86, with $T_\mathrm{eff}$ = 4730 K, from the {Kepler Input Catalogue \citepads{2011AJ....142..112B}}, versus KIC8462's $T_\mathrm{eff}$ = 6750 K, from B15) was included in the same entrance slit and used as reference in the photometric analysis.

\section{Extraction of  multicolour photometry}
The reduction procedure follows the precepts of \citetads{2016A&A...589L...6A}, with the difference that a zero-order spectrum is not recorded while using the R1000R grism. Instead, the shifts of the spectra in the wavelength region were estimated and corrected a-posteriori using the H$_\alpha$ line for both the target and comparison. An aperture of 44 pixels was used to extract the flux from the spectra, and the sky-background was estimated on regions close to the spectra. This relatively wide aperture was needed as the spectra were slightly defocussed, both to avoid saturation of the reference star and to try to minimize the errors due to inefficient flat-field correction.  

The extracted spectra were divided into five wavelength-bands that avoid significant telluric and stellar lines. The selected regions are presented in Fig.~\ref{fig:WTFspectrum}. Final fluxes in each band were obtained from the average flux during the time-series acquired at each of the pointings.  The corresponding flux-errors are the uncertainties of the mean of the time-series, that is, their standard-deviation divided by $\sqrt N$, were N is the number of individual measures in a given time-series. The fluxes and their errors were then normalised to the average flux from the pointings Nr. 6, 7, 15 and 16, when KIC8462 had a brightness close to the one before the first dimming event. This normalisation procedure was performed independently for each of the bands.  Table~\ref{tbl:fluxes} indicates these averaged and normalised fluxes for each pointing, while Fig.~\ref{fig:lco_gtc_phot} displays them together with nightly photometry obtained by two 0.4m telescopes of the LCO (Las Cumbres Observatory) network. Two GTC pointings, Nr. 1 and 12, were rejected in the further analysis since their colours  presented a much larger scatter than the others{\footnote{The inclusion of either of the rejected pointings increased the $\chi^2$ of the fitting reported in Section 4 by more than 40\%}. This was due to inconsistent  exposure times during the first pointing, and poor sky conditions at the twelfth pointing. As is evident, the bluer wavelength-bands from the GTC data undergo deeper dimming events (see also Fig.~\ref{fig:narrow_vs_depth}); this is analysed in the next section. The LCO data are being described in the parallel publication about the community follow-up effort (B18).

\begin{table*}
\centering
\caption{Observed Fluxes at each pointing in five wavelengths bands, indicated by their central wavelengths in $nm$. The BJD times refer to the central time of each pointing.}
\begin{tabular}{llcccccc}
\hline
\hline
Nr&BJD-2400000 & $F_{556}$&$F_{618}$&$F_{709}$&$F_{792}$&$F_{868}$&\\
\hline

\ 1\tablefootmark{a}&57890.6940&0.9942$\pm$0.0020&0.9984$\pm$0.0010&1.0046$\pm$0.0011&1.0010$\pm$0.0010&0.9986$\pm$0.0012& \\
\ 2&57891.6709&0.9917$\pm$0.0018&0.9946$\pm$0.0018&0.9976$\pm$0.0028&0.9971$\pm$0.0027&0.9961$\pm$0.0013& \\
\ 3&57893.6921&0.9879$\pm$0.0008&0.9901$\pm$0.0008&0.9936$\pm$0.0010&0.9948$\pm$0.0010&0.9943$\pm$0.0009& \\
\ 4&57895.6698&0.9933$\pm$0.0010&0.9951$\pm$0.0008&0.9961$\pm$0.0011&0.9967$\pm$0.0007&0.9960$\pm$0.0007& \\
\ 5&57896.6644&0.9950$\pm$0.0013&0.9963$\pm$0.0009&0.9954$\pm$0.0012&0.9954$\pm$0.0009&0.9971$\pm$0.0009& \\
\ 6&57897.6662&0.9976$\pm$0.0013&1.0002$\pm$0.0010&1.0002$\pm$0.0015&1.0011$\pm$0.0011&1.0013$\pm$0.0013& \\
\ 7&57902.7096&0.9969$\pm$0.0015&0.9994$\pm$0.0010&1.0004$\pm$0.0018&0.9994$\pm$0.0017&1.0005$\pm$0.0014& \\
\ 8&57921.6949&0.9813$\pm$0.0009&0.9861$\pm$0.0007&0.9912$\pm$0.0012&0.9940$\pm$0.0011&0.9933$\pm$0.0013& \\
\ 9&57922.5738&0.9898$\pm$0.0013&0.9926$\pm$0.0009&0.9941$\pm$0.0015&0.9957$\pm$0.0013&0.9956$\pm$0.0009& \\
10&57980.5654&0.9938$\pm$0.0010&0.9945$\pm$0.0015&0.9945$\pm$0.0021&0.9961$\pm$0.0021&0.9970$\pm$0.0015& \\
11&57985.6109&0.9995$\pm$0.0025&1.0008$\pm$0.0020&1.0038$\pm$0.0018&1.0026$\pm$0.0012&1.0007$\pm$0.0019& \\
12\tablefootmark{b}&57986.6427&0.9991$\pm$0.0015&1.0043$\pm$0.0014&1.0129$\pm$0.0019&1.0063$\pm$0.0009&1.0017$\pm$0.0012& \\
13&58007.3854&0.9780$\pm$0.0013&0.9811$\pm$0.0008&0.9855$\pm$0.0013&0.9882$\pm$0.0010&0.9879$\pm$0.0014& \\
14&58091.3543&1.0051$\pm$0.0010&1.0026$\pm$0.0015&1.0021$\pm$0.0021&1.0010$\pm$0.0027&0.9998$\pm$0.0014& \\
15&58092.3331&1.0044$\pm$0.0015&1.0007$\pm$0.0016&0.9996$\pm$0.0010&0.9994$\pm$0.0022&0.9988$\pm$0.0014& \\
16&58093.3502&1.0011$\pm$0.0013&0.9997$\pm$0.0013&0.9997$\pm$0.0015&1.0001$\pm$0.0019&0.9994$\pm$0.0018& \\

\hline
\hline
\end{tabular}
\tablefoot{
\tablefoottext{a}{Data not used in analysis due to large scatter from varying exposure times}
\tablefoottext{b}{Data not used due to poor sky transparency}
}
\label{tbl:fluxes}
\end{table*}

\begin{figure}
   \centering
   \includegraphics[width=9.cm]{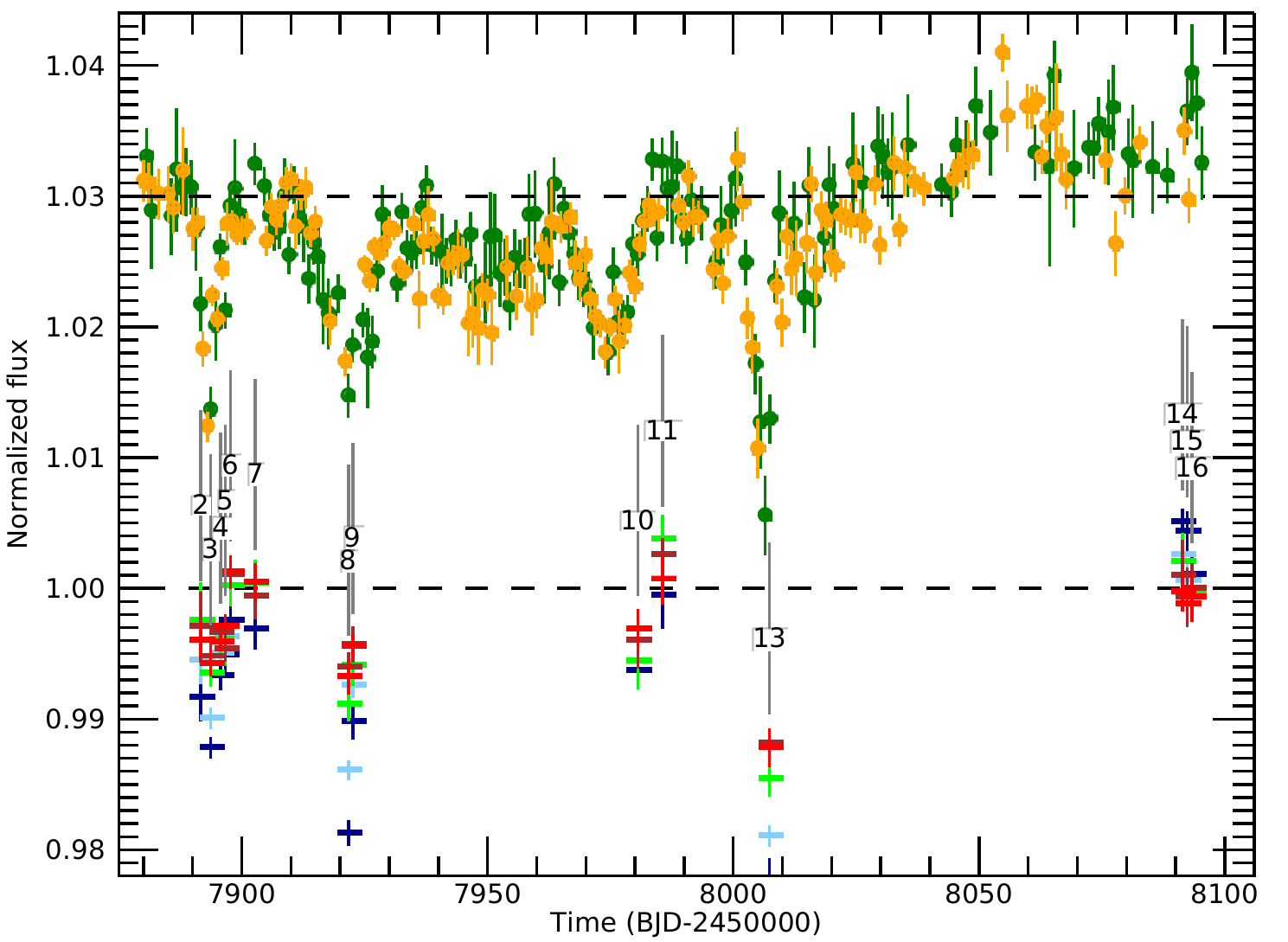}
  
   \caption{
   Normalized fluxes of KIC 8462852 from 6th of May to 8th of December 2017, from LCO and GTC. The upper light curve shows nightly photometry in the r' filter from the 0.4m LCO telescopes in Tenerife (green) and Hawaii (orange), upwards shifted by 0.03 flux units for a clearer display. The assumed quiescent flux is indicated by a dashed line. The lower symbols show the corresponding GTC narrow-band fluxes with their errors. The crosses' colours correspond to the wavelength ranges indicated by Fig. ~\ref{fig:WTFspectrum}. The vertical lines indicate the time of each numbered pointing. Only the 14 pointings used in the analysis are shown. }

     \label{fig:lco_gtc_phot}
    \end{figure}

 \begin{figure}
   \centering
   \includegraphics[width=9.cm]{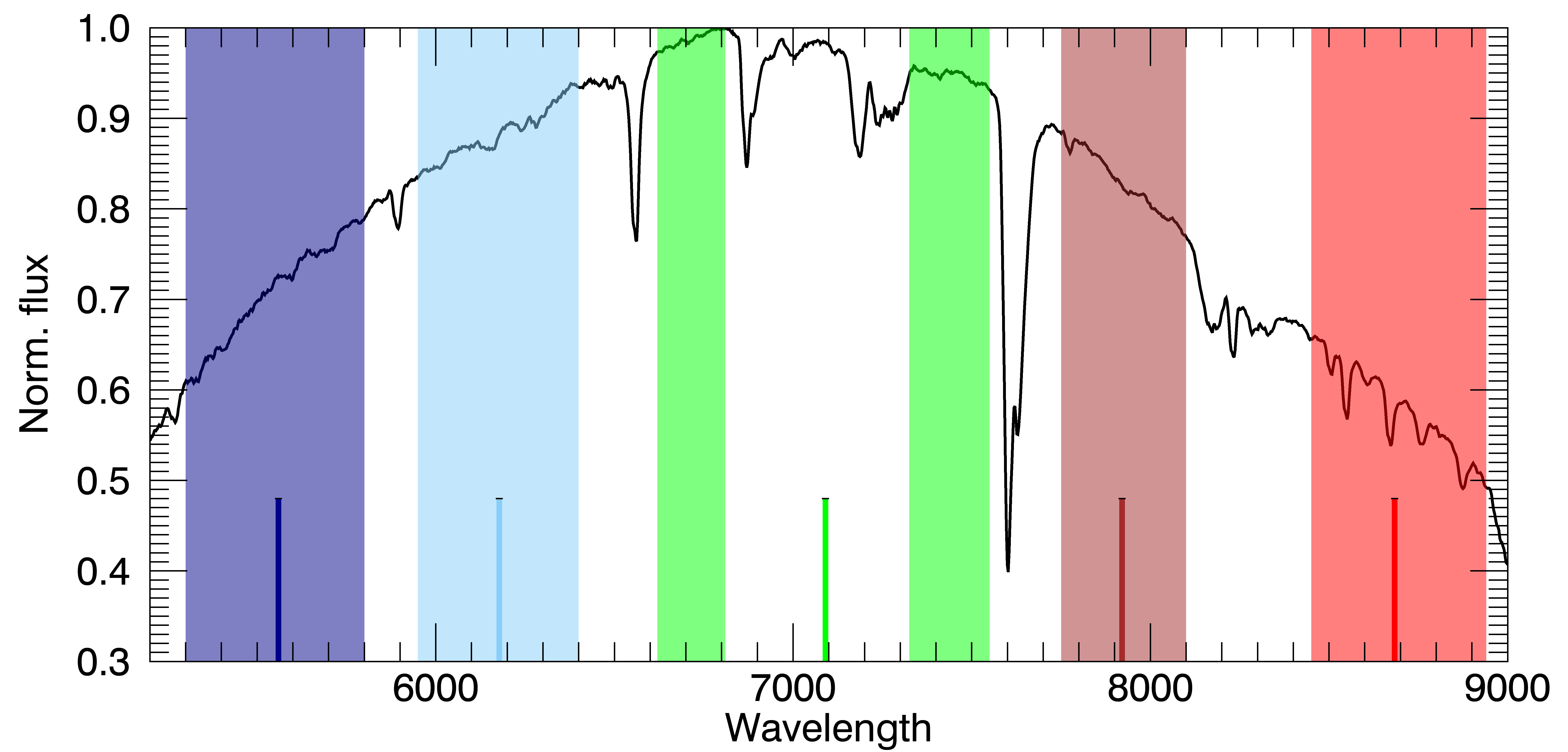}  
   \caption{KIC 8462852's spectrum, with the featureless ranges that were extracted for photometry. The bands' flux-weighted central wavelengths are: blue: 556nm, light blue: 618nm, lime: 709 nm, brown: 792nm, red: 868nm; they are indicated by the vertical lines at the bottom of the plot.}
     \label{fig:WTFspectrum}
    \end{figure}

\section{Assessment of the spectrophotometry}
The analysis of the wavelength-dependent photometric data is also building on the methodology of \citetads{2016A&A...589L...6A}, who analysed a similar data-set for the presence of colour-signatures. In brief, an optically thin occulter of uniform optical depth is assumed to be responsible for the dimming events, which covers a fraction $f$ of the surface of the star. The fractional depth of a dimming-event is then given by $D_n = \tau_n \,f ~ $, where $D_n$ and $\tau_n$ refers to the depth and optical depth, respectively, at the wavelength $\lambda_n$. The occulter is assumed to consist of dust-like particles with an extinction cross-section  $\sigma_{ext}$ for which a wavelength dependency of the form $\sigma_{ext} \propto \tau \propto \lambda^{-\alpha}$ is assumed, leading to:
 
 \begin{equation}
D_n \propto (\lambda_n)^{-\alpha}
\label{eq:Dn}
\end{equation}
 
  For a dust column of fixed density and composition, this defines the exponent $\alpha$, also known as the Absorption \ang Coefficient (AAC) \citep{moosmuller11}. The use of the AAC  is advantageous over the quoting of extinction or opacity ratios between different wavelengths, since \ang coefficients \citep{angstrom29} do not have a first-order dependency on the wavelength-ratio, as do these ratios.

 Similar spectrophotometry taken by GTC is analyzed  by \citeads[][]{2016A&A...589L...6A}, who derive their  \ang coefficient from the slope of measures of $D$ at two different wavelengths (see their Eqs. 3 - 5). Re-enacting this procedure with the data of KIC8462, the $\alpha$ values were however varying strongly in dependence on the wavelengths being paired, and an $\alpha$ value common to our five bands was very sensitive to each pairing's weight. Therefore, in this work we use a new procedure that derives a common value of $\alpha$ from a simultaneous fit to the acquired depths at all bands, as follows.

Using the relation between depth and wavelength of Eq.~\ref{eq:Dn}, we assume that the modelled depths $D_{mod,n,i}$ at the wavelengths $\lambda_n$ relate to reference-depths  $D_{w,i}$ at the wavelength $\lambda_w$ with: 
\begin{equation}
 \frac{D_{mod,n,i}+\Delta}{D_{w,i}+\Delta} = \left(\frac{\lambda_n}{\lambda_w}\right)^{-\alpha} 
 \label{eq:fitfct}
 \end{equation}
where the indices $n$ indicate the bands and $i$ stands for each telescope pointing. $\Delta$ is an offset common to all wavelengths, which accounts for any uncertainties in the global flux-normalisation.
The reference values $D_{w,i}$ were obtained for
each pointing $i$ from the average of the Depths of all the five bands. For the model fitting 
function, Eq.~\ref{eq:fitfct} was resolved for $D_{mod,n,i}$ and a $\chi^2$ minimisation was performed for:
 \begin{equation}
 \chi^2 = \sum{\frac{(D_{mod,n,i}(\alpha, \Delta) - D_{n,i})^2}{\sigma_{D_{n,i}}^2}}
  \end{equation}
 with {three} free parameters to be fitted, a namely $\alpha$, $\Delta$ {and $\lambda_w$. The initial value for the reference wavelength $\lambda_w$ was the flux-weighted mean from all five band-passes, at $709 nm$. Including $\lambda_w$ a as a free parameter led to a marked improvement in the fit, as shifts in $\lambda_w$ may account for the weighting of the individual bands' fluxes when obtaining $D_w$.}
 From an AMOEBA minimisation of above equation, a best fitting value for the AAC was {obtained as 
\begin{equation}
\alpha =  2.19\pm0.45
\end{equation}
with an offset of $\Delta = -0.0009\pm 00005$ in normalized flux units, and a wavelength $\lambda_w = 682\pm 15 \mu m$. }The relation between the depths $D_{n,i}$ in the various wavelengths versus the reference depths $D_{w,i}$ is shown in Fig.~\ref{fig:narrow_vs_depth}, together with the best-fitting model values $D_{mod,n}(D_w)$ for each wavelength $\lambda_n$. In this figure, $\Delta$ is visible as the amount by which the crossing among the model lines is shifted against the zero-points of the $D_{n,i}$ and $D_{w,i}$, whereas changes in the reference wavelength $\lambda_w$ correspond to a change in the gradient of the entire set of model lines.
  
We noted already that the quoted errors of the $D_{n,i}$ are the error of the mean of the time-series that led to each flux-value. Besides these errors, we expect difficult-to-quantify errors from the flux-calibration of the target against the reference star, and errors from other sources, such as flat fielding errors or effects from differential atmospheric transmission. The model with the best-fitting values of $\alpha$ and $\Delta$ had a reduced Chi-square of {3.1}. If we assume that the fitted model is correct, a reduced chi-square of $\approx$1 should have been obtained, however. In order to achieve this, a second component would have to added to the quoted flux-errors, which is about 1.5 times larger than these. 
{This second component, as well as a variation  of $\alpha$ of $\approx 0.15$, observed if different sets of pointings were used for the flux-normalization of Sect. 3, has been included in the quoted error of $\alpha$ (Eq. 4). We also consider in the quoted error the effects of considering only subsets of the data: Using only the even numbered points rises $\alpha$ by 0.5, whereas the odd points on their own lower it by 0.4. 

We also note that the red-most band in Fig.~\ref{fig:narrow_vs_depth} has a poorer fit than the other bands. Omitting that band from the analysis, the AAC across the remaining four bands increases to $\alpha = 2.30$.
}
 
\begin{figure}
   \centering
    \includegraphics[width=9.cm]{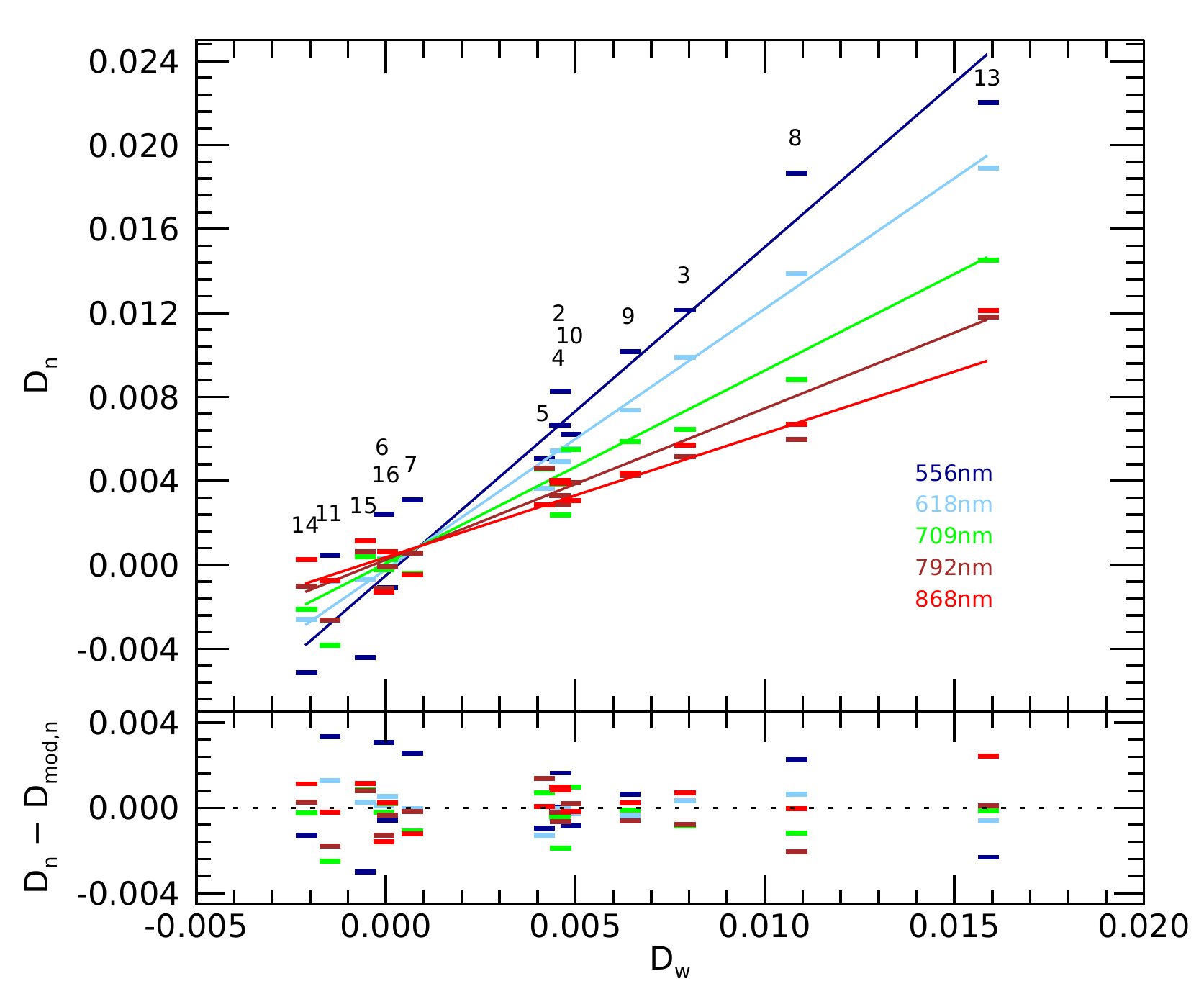}
  
   \caption{Plot of the measured depths $D_n$ in the five colour-bands, indicated by the same colours as in Fig~\ref{fig:WTFspectrum}, versus the white (average) depths $D_w$, for the 14 pointings that were analysed. The straight lines indicate the joint-fit to all bands, with the best common value for the AAC of $\alpha = 2.19$. The lower panel shows the residuals, on the same scale. Error bars have been omitted for clarity. 
   }
     \label{fig:narrow_vs_depth}
    \end{figure}

 \begin{figure}
   \centering
   \includegraphics[width=9.cm]{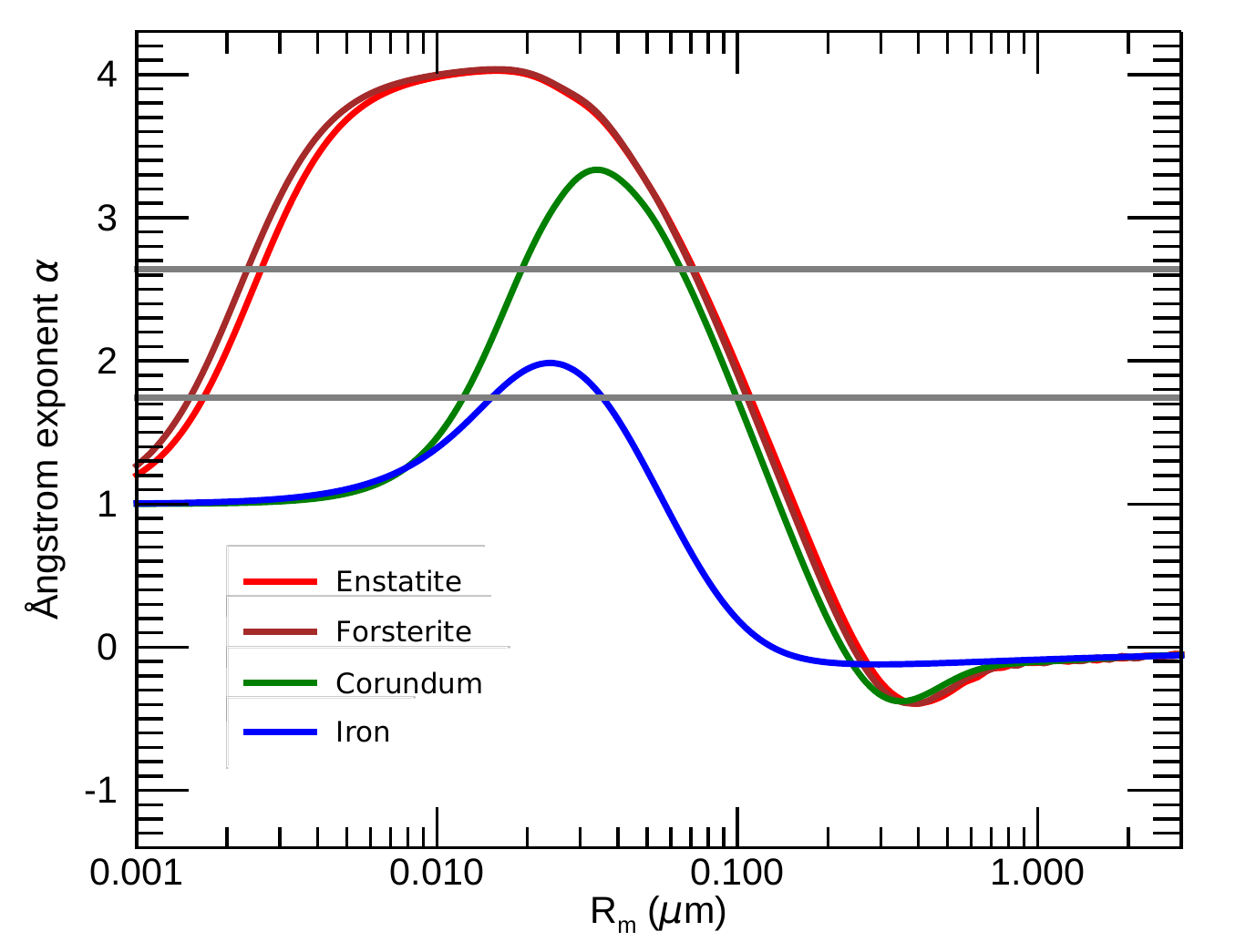}
  
   \caption{Calculated absorption \ang exponents $\alpha$ as a function of particle size for four different minerals. For each mineral we used a log-normal particle size distribution, where $R_m$ is the median particle size in the distribution. The two grey horizontal lines denote the $\pm 1\sigma$ bounds to the \ang exponent obtained from the GTC observations.
}
     \label{fig:materials}
    \end{figure}

\section{Discussion: Absorbers that may cause the dimming events}
The dimming events of KIC8462 that have been observed, of depths of up to 1.6\%, have a clear colour-signature, with the dips approximately 1.5 - 2 times deeper in the bluer than in the redder bands. This dependency is analysed in terms of the dimmings' absorption \ang coefficient (AAC), found to be $\alpha = 2.19\pm 0.45$, which describes the wavelength-dependency of the opacity of an absorbing material, assumed to be optically thin.

In Fig.~\ref{fig:materials} we show the AACs for several potential absorbing materials. These coefficients are based on Mie extinction cross-sections\footnote{Cross-sections for log normal distributions were calculated with the set of Mie scattering  routines made available by the Earth Observations Data Group of the University of Oxford, at \url{http://eodg.atm.ox.ac.uk/MIE/}}, for spherical particles with log-normal size distributions around the indicated median sizes $R_m$, with the real and imaginary refraction indices given in \citetads[][{ see also for a plot assuming narrower size distributions of the particles}]{2016A&A...589L...6A}. The plotted AACs were then derived from the materials' cross sections at  587 and 830 $nm$, which are averages between our two bluest and reddest bands, respectively. 
As we can see in Fig.~\ref{fig:materials}, the range of $\alpha$ from the dimming events is compatible with several combinations of particle sizes and materials. Given the uncertainties in the true nature of the particles' composition, shapes and size-distribtion, we may only conclude that an optically thin absorber must be dominated by particles that are smaller than $\approx 0.15  \mu m$. In the Raleigh regime, when the particle size is much smaller than the wavelength, the AAC tends towards 1 \citep{moosmuller11}, and sizes less than $\approx 0.0015 \mu m$ may also be excluded with reasonable certainty. An optically thin absorber must therefore be dominated by particles with sizes in the Mie-regime.

We have to emphasize that the quoted value for the AAC comes from different dimming events, and dust properties may be different among them. It is of note that the measure at the deepest dip (Nr. 13) has less spread in colour than the second deepest one (Nr. 8, see Fig.~\ref{fig:narrow_vs_depth}). {If point Nr. 13 is discarded, the AAC rises by  0.38.  Conversely, if Nr. 8 is discarded, the AAC gets lowered by 0.29.}. 

The observed AAC has been derived under the assumption of an optically thin absorber. It may however well be the case that at least part of the absorbing cross-section is optically thick. In this case, the optically thin cross-section would need to have a higher AAC, in order to compensate for the optically thick sections that on their own, have a value of $\alpha = 0$. The observed value of $\alpha = 2.19\pm0.45$ is therefore a lower limit. However, as we see in Fig.~\ref{fig:materials}, an increase of $\alpha$ by 1 - 1.5 does not generate significant new constrains onto the absorbing materials, with the exception that iron, due to its large refractive index, may be excluded. 

The observed colour-dependency of KIC8462's dimming does therefore support, but not prove, the hypothesis that occultation by dusty material is the principal origin of the dimming events. Such occulters, in the form of a few distinct compact planetesimals envolved in a  wide dust-clouds have been proposed by \citetads{2017A&A...600A..86N} 
 to be responsible for the dimming events that were observed by Kepler.

{Blow-out of dust particles will occur if radiation pressure from the central star becomes a significant fraction of the gravity that determines the particles orbit. This fraction, known as the $\beta$ parameter \citepads{1979Icar...40....1B}, surpasses a value of $\beta=0.5$ for particles of $\la 1\mu$m  \cite[see Fig. 2 of ][]{2017A&A...600A..86N}. Longer-term brightness variations -- if they are caused by dust particles -- may therefore only arise from dust with significantly larger particles. Following this argumentation, \citetads{2017arXiv171005929W} predicted that such variations should be of neutral colour at visible wavelenghts (see their Fig. 3).   

 Indeed,  
 \citetads{2017ApJ...847..131M} 
 report a gradual fading between October 2015 and December 2016 
with a relatively neutral colour. They quote it in terms of the extinction ratio $R_V$ \citepads{1989ApJ...345..245C}
, with $R_V \ga 5$. In terms of the AAC, this corresponds\footnote{We base this conversion on the fact that ratios of absolute extinctions $A_n / A_m$ are identical to the depth ratios used in our work; that is, $A_n / A_m = D_n / D_m$ as long as the $A$ and $D$ are $<< 1$, with $n$ and $m$ representing two wavelenghts.  \citetads{1989ApJ...345..245C} use $A(\lambda)/A_V$ a function of $R_V$ to describe colour-dependence of interstellar extinction through a single number $R_V$. For a given $R_v$ and a set of wavelengts $\lambda_i$, we may then calculate all values $A(\lambda_i)/A_V$ using their Eq. 3. Through $\chi^2$ minimization we may then find the value $\alpha$ that generates the set of $D(\lambda_i)/D_V = (\lambda_i/ \lambda_V )^{-\alpha}$ that most closely matches the set $A(\lambda_i)/A_V$. This procedure can easily be reversed to determine $R_V$ for a given $\alpha$. We note that both $R_V$ and the AAC generate phenomenological descriptions of wavelength dependencies without implying underlying physical models.} to a value of $\alpha \la 1.1$. {For the long-term fading during the Kepler mission \citepads{2016ApJ...830L..39M}
, a similar value of $R_V = 5.0 \pm 0.9$ (or $\alpha = 1.1 \pm 0.1$) was also quoted by \citetads{2017arXiv171204948D}
, based on NUV photometry acquired in 2011 and 2012 by the GALEX mission, which they compared to Kepler photometry.}
The long-term fading may therefore have much less colour-dependence than the short events observed by us and by B18. However, the colours by \citetads{2017ApJ...847..131M} are derived from only two epochs, comparing fluxes before and after an observational gap in winter 2015/16, {while the colours by \citetads{2017arXiv171204948D} are based on a wavelength interval rather different from the one considered in this work}. We consider it therefore too early to ascertain that the longer-term flux variations have a fundamentally different origin (or very different absorbers) than the short dimming events. Continued longer term coverage of KIC8462 in multiple wavelengths is certainly desirable to clarify if a difference exists between the colours of the dimming events, on time-scales of days, and the variations on time-scales of months and longer.

Considering again the short dimming events, \citetads{2017arXiv171005929W} infer from the duration of the deepest ones in the Kepler data, that any occulters that might cause them} have to be on highly eccentric orbits with pericenters at < 0.6AU. This would imply still larger sizes for dust particles, in order to become resistent to blow-out. Our data, and as well those by B18, stipulate however that the dimmings are generated by smaller dust particles, with $R_m \la 0.15 \mu$ (this work) or $R_m \la 0.2 \mu$ (B18). 

 For such fine dust, radiation pressure is larger than gravitational attraction by the star ($\beta > 1$). Dust with only weak gravitational bounding to an orbiting body will therefore decouple from it and move on a hyperbolic trajectory away from the star and the orbiter \citep[see Fig. 3 of ][]{2017A&A...600A..86N}. This decoupling is rather fast: At a distance of 0.6AU from KIC8462, acceleration from radiation pressure will move dust with $\beta = 1$ within about 3.5 days over a distance of KIC8462's radius (1.58R$\sun$, B18). If we assume a dust size of 0.1 $\mu m$, then $\beta$ may take values of 2-10, depending on the material, and the time to cover the same  distance will be reduced by a factor of $\beta^{-1/2}$. 
The presence of such fine dust during KIC8462's short dimming events requires therefore that this dust is constantly being replenished during the occulter's orbit. We can therefore exclude an occulter that consists only of dust, but rather, require a central object that is able to regenerate that dust during the flyby on KIC8462. }

If dust is indeed the source of dips in KIC8462, it is also to be expected that the observation of much deeper dips like those found in the Kepler data will show that these have less of a colour signature. The reason is that a dimming depth of 15-20\% is difficult to generate by an absorber that is optically thin across the {\it entire} occulted stellar surface. In order to {\it not} be optically thick, such a strong absorber would have to be of rather uniform optical thickness, with values on the order of $\tau \approx 0.1 $ on a size-scale of the stellar diameter. More likely, the absorber has significant non-uniformities and at least part of it will be optically thick during the deeper events.

The observational coverage of a deep event with simultaneous multicolour photometry of with spectrophotometry, and preferentially with a dense temporal coverage, should therefore be central to resolve the hypothesis of a mostly dusty absorber as the source of the dimming events on KIC 8462852.

\section{Summary}
{The principal results of this work are summarised below:

- In dimming events of KIC8462 of depths of up to 1.6\% and lasting a few days, we find a clear colour-signature, with the dips approximately 1.5 - 2 times deeper in the bluer than in the redder visual bands. The wavelength dependency can be described by a single number, an \ang Absorption Coefficient (AAC) of 2.19$\pm$0.45. 

- This is different to KIC8462’s flux variations on time-scales of months to years, which have been reported to display largely neutral colours in the visual regime. These variations should therefore be caused by different absorbers, or arise from different processes, than the short events.

- Assuming that the short events are caused by occultations, their wavelength dependency implies that most of the cross-section of the occulters must be optically thin, and consist of particles with sizes in the range 0.0015 – 0.15 $\mu m$.

- Such small particles will not resist blow-out by radiation pressure. At expected periastron distances of 0.6AU or less, radiation pressure will deviate them on time-scales of days over distances larger than then radius of the central star. They must therefore be replenished continually.

- In the 15-20\% deep events observed to date only by Kepler, a significant part of their occulting cross section was likely optically thick. Hence we predict for deeper events a tendency towards more neutral colour signatures. This tendency may already be present in the deepest event that was observed by GTC.
}

\section{Acknowledgements}
{We thank the referee, Benjamin T. Montet of the California Institute of Technology, for the rapid and thorough revision, which led to a marked improvement in the presentation of this work.} This research made use of data acquired with the Gran Telescopio Canarias (GTC), installed at the Spanish Observatorio del Roque de los Muchachos of the Instituto de Astrof\'\i sica de Canarias, in the island of La Palma. HD, RA and DNe acknowledge support by grants ESP2015-65712-C5-4-R, ESP2016-80435-C2-2-R, and RYC-2010-06519 of the Spanish Secretary of State for R\&D\&i (MINECO).

\bibliographystyle{aa} 
\bibliography{WTF1}

\begin{thebibliography}{16}
\expandafter\ifx\csname natexlab\endcsname\relax\def\natexlab#1{#1}\fi

\bibitem[{{Alonso} {et~al.}(2016){Alonso}, {Rappaport}, {Deeg}, \&
  {Palle}}]{2016A&A...589L...6A}
{Alonso}, R., {Rappaport}, S., {Deeg}, H.~J., \& {Palle}, E. 2016, \aap, 589,
  L6

\bibitem[{{\AA ngstr\"om}(1929)}]{angstrom29}
{\AA ngstr\"om}, A. 1929, Geogr. Ann., 11, 156

\bibitem[{{Bodman} \& {Quillen}(2016)}]{2016ApJ...819L..34B}
{Bodman}, E.~H.~L. \& {Quillen}, A. 2016, \apjl, 819, L34

\bibitem[{{Boyajian} \& {et al.}(2018)}]{B18}
{Boyajian}, T.~S. \& {et al.} 2018, accepted for ApJL, ArXiv e-prints (B18)

\bibitem[{{Boyajian} {et~al.}(2016){Boyajian}, {LaCourse}, {Rappaport},
  {Fabrycky}, {Fischer}, {Gandolfi}, {Kennedy}, {Korhonen}, {Liu}, {Moor},
  {Olah}, {Vida}, {Wyatt}, {Best}, {Brewer}, {Ciesla}, {Cs{\'a}k}, {Deeg},
  {Dupuy}, {Handler}, {Heng}, {Howell}, {Ishikawa}, {Kov{\'a}cs}, {Kozakis},
  {Kriskovics}, {Lehtinen}, {Lintott}, {Lynn}, {Nespral}, {Nikbakhsh},
  {Schawinski}, {Schmitt}, {Smith}, {Szabo}, {Szabo}, {Viuho}, {Wang},
  {Weiksnar}, {Bosch}, {Connors}, {Goodman}, {Green}, {Hoekstra}, {Jebson},
  {Jek}, {Omohundro}, {Schwengeler}, \& {Szewczyk}}]{2016MNRAS.457.3988B}
{Boyajian}, T.~S., {LaCourse}, D.~M., {Rappaport}, S.~A., {et~al.} 2016,
  \mnras, 457, 3988

\bibitem[{{Brown} {et~al.}(2011){Brown}, {Latham}, {Everett}, \&
  {Esquerdo}}]{2011AJ....142..112B}
{Brown}, T.~M., {Latham}, D.~W., {Everett}, M.~E., \& {Esquerdo}, G.~A. 2011,
  \aj, 142, 112

\bibitem[{{Burns} {et~al.}(1979){Burns}, {Lamy}, \&
  {Soter}}]{1979Icar...40....1B}
{Burns}, J.~A., {Lamy}, P.~L., \& {Soter}, S. 1979, \icarus, 40, 1

\bibitem[{{Cardelli} {et~al.}(1989){Cardelli}, {Clayton}, \&
  {Mathis}}]{1989ApJ...345..245C}
{Cardelli}, J.~A., {Clayton}, G.~C., \& {Mathis}, J.~S. 1989, \apj, 345, 245

\bibitem[{{Davenport} {et~al.}(2017){Davenport}, {Covey}, {Clarke}, {Laycock},
  {Fleming}, {Boyajian}, {Montet}, {Shiao}, {Million}, {Wilson}, {Olmedo},
  {Mamajek}, {Olmedo}, {Chavez}, \& {Bertone}}]{2017arXiv171204948D}
{Davenport}, J.~R.~A., {Covey}, K.~R., {Clarke}, R.~W., {et~al.} 2017, ArXiv
  e-prints [\eprint[arXiv]{1712.04948}]

\bibitem[{{Meng} {et~al.}(2017){Meng}, {Rieke}, {Dubois}, {Kennedy}, {Marengo},
  {Siegel}, {Su}, {Trueba}, {Wyatt}, {Boyajian}, {Lisse}, {Logie}, {Rau}, \&
  {Vanaverbeke}}]{2017ApJ...847..131M}
{Meng}, H.~Y.~A., {Rieke}, G., {Dubois}, F., {et~al.} 2017, \apj, 847, 131

\bibitem[{{Montet} \& {Simon}(2016)}]{2016ApJ...830L..39M}
{Montet}, B.~T. \& {Simon}, J.~D. 2016, \apjl, 830, L39

\bibitem[{{Moosm\"uller}(2011)}]{moosmuller11}
{Moosm\"uller}, H. 2011, Atmos.Chem. Phys., 11, 10677

\bibitem[{{Neslu{\v s}an} \& {Budaj}(2017)}]{2017A&A...600A..86N}
{Neslu{\v s}an}, L. \& {Budaj}, J. 2017, \aap, 600, A86

\bibitem[{{Schaefer}(2016)}]{2016ApJ...822L..34S}
{Schaefer}, B.~E. 2016, \apjl, 822, L34

\bibitem[{{Wright} \& {Sigurdsson}(2016)}]{2016ApJ...829L...3W}
{Wright}, J.~T. \& {Sigurdsson}, S. 2016, \apjl, 829, L3

\bibitem[{{Wyatt} {et~al.}(2018){Wyatt}, {van Lieshout}, {Kennedy}, \&
  {Boyajian}}]{2017arXiv171005929W}
{Wyatt}, M.~C., {van Lieshout}, R., {Kennedy}, G.~M., \& {Boyajian}, T.~S.
  2018, \mnras, 473, 5286

\end{thebibliography}

\end{document}